\documentclass[useAMS,usenatbib]{mnras}
\usepackage{graphicx}
\usepackage{aas_macros}

\title[BeXRBs in SMC]{Confirmation of six Be X-ray binaries in the Small Magellanic Cloud\thanks{Based on 
ESO data from 079.D-0371, 088.D-0352}}

\author[V.A. McBride et al.]
{\parbox{\textwidth}{V.A.~McBride$^{1,2}$\thanks{E-mail: \texttt{vanessa@saao.ac.za} (VAM)}, A.~Gonz\'{a}lez-Gal\'{a}n$^3$, A.J.~Bird$^{4}$, M.J.~Coe$^{4}$, E.S.~Bartlett$^{1,5}$, R.~Dorda$^{3}$, F.~Haberl$^{5}$, A.~Marco$^{3}$, I.~Negueruela$^{3}$, M.P.E.~Schurch$^{1}$, R.~Sturm$^{5}$, D.A.H.~Buckley$^{2}$ and A.~Udalski$^{6}$}\vspace{0.4cm}\\
$^{1}$Department of Astronomy, University of Cape Town, Private Bag X3, Rondebosch, 7701, South Africa\\
$^{2}$South African Astronomical Observatory, PO Box 9, Observatory, 7935, South Africa\\
$^{3}$Departamento de F\'isica, Ingenier\'ia de Sistemas y Teor\'ia de la Se\~nal, Universidad de Alicante, Apdo. 99, E03080 Alicante, Spain\\
$^{4}$School of Physics and Astronomy, University of Southampton, Highfield, Southampton SO17 1BJ, United Kingdom\\
$^{5}$ESO - European Southern Observatory, Alonso de C\'{o}rdova 3107, Vitacura, Casilla 19001, Santiago de Chile, Chile\\
$^{6}$Max-Planck-Institut f\"ur extraterrestrische Physik, Giessenbachstra\ss e, 85748 Garching, Germany\\
$^{7}$Warsaw University Observatory, Aleje Ujazdowskie 4, 00-478, Warsaw, Poland
}
\begin{document}

\date{Accepted 2017 January 19. Received 2017 January 18 ; in original form 2016 July 22}

\pagerange{\pageref{firstpage}--\pageref{lastpage}} \pubyear{2002}

\maketitle

\label{firstpage}

\begin{abstract}
The X-ray binary population of the Small Magellanic Cloud (SMC) contains a large number of massive X-ray binaries and the recent survey of the SMC by {\it XMM-Newton} has resulted in almost 50 more tentative High Mass X-ray Binary (HMXB) candidates. Using probability parameters from \citet{HaberlSturm2016} together with the optical spectra and timing in this work, we confirm six new massive X-ray binaries in the SMC. We also report two very probable binary periods; of 36.4\,d in XMM\,1859 and of 72.2\,d in XMM\, 2300. These Be X-ray binaries are likely part of the general SMC population which rarely undergoes an X-ray outburst.
\end{abstract}


\section{Introduction}

The Small Magellanic Cloud (SMC) has a unique population of X-ray binaries.  It is entirely dominated by massive X-ray binaries, all of which have neutron star companions. This is a factor of $\sim$50 more X-ray binaries than would be expected, based on scaling by mass the Milky Way X-ray binary population. This excess is thought to be due to the higher star formation rate in the SMC, but possibly also influenced by the lower metallicity compared to the Milky Way \citep{Dray2006}.  The massive X-ray binary populations in the Milky Way, Large Magellanic Cloud (LMC) and SMC all have donor stars with spectral types predominantly earlier than B3 \citep{Negueruela1998,NegueruelaCoe2002, McBrideCoeNegueruela2008,AntoniouHatzidimitriouZezas2009, AntoniouZezas2016}.  Our current understanding of this limited mass range is through angular momentum loss during evolution of the binary \citep{PortegiesZwart1995}. 

With a view to characterising the X-ray population of the SMC, \citet{HaberlSturmBallet2012} have undertaken an X-ray survey of the SMC down to a limiting luminosity of $5\times10^{33}$\,erg\,s$^{-1}$ (assuming a distance of 60\,kpc to the SMC, \citealt{HilditchHowarthHarries2005}), with wide coverage across the SMC (see Fig.~3 in \citealt{HaberlSturmBallet2012}). Through the primary analysis of the point source X-ray data from this survey, \citet{SturmHaberlPietsch2013} have generated a list of candidate high mass X-ray binaries (HMXBs) which is presented in Table 5 of their paper. Objects are classified as candidate HMXBs on the basis of their X-ray and optical colours. In particular, hard X-ray sources are classified as HMXBs when they have an optical counterpart in the magnitude range $13.5<V<17$ with optical colours characteristic of an early-type star, and are not already identified with a known background AGN.  Forty-five X-ray point sources are classified as candidate HMXBs by \citet{SturmHaberlPietsch2013}.

In this paper we selected seven of the most likely HMXB candidates from the X-ray classification by \citet{SturmHaberlPietsch2013} for follow-up optical spectral and temporal analysis. From this analysis we confirm these candidates as Be X-ray binaries in the SMC. In section \ref{Data} we present the photometric and spectroscopic data used in the analysis, while in section \ref{Method} we discuss the selection criteria and analysis techniques.  Our results are presented in section \ref{Results}, while conclusions are discussed in section \ref{Conclude}.

\section[]{Data}
\label{Data}

\subsection{Spectroscopy from ESO}

Spectra were taken between 2011 December 9 and 10 with the ESO Faint Object Spectrograph (EFOSC2) mounted at the Nasmyth B focus of the 3.58\,m NTT. A slit width of $1.5^{\prime\prime}$ was employed, together with a grating ruled at $600$\,l\,mm$^{-1}$ that yielded $1$\,\AA /pixel dispersion over a wavelength range of $\lambda\lambda3095-5085$\,\AA . 
Spectra were recorded with exposure times between 300\,s and 800\,s depending on source brightness, at a spectral resolution of $\sim10$\AA\,.

We also utilised archival spectra of two objects observed on 2007 September 20 with EFOSC2 mounted on the ESO 3.6\,m telescope at La Silla. In this instance a slit width of $1.0^{\prime\prime}$ was employed with all other instrument parameters the same as above. Typical exposure times ranged between 1000\,s -- 1500\,s.

All data were reduced using the standard packages available in the Image Reduction and Analysis Facility (IRAF\footnote{IRAF is distributed by the National Optical Astronomy Observatory, which is operated by the Association of Universities for Research in Astronomy (AURA) under cooperative agreement with the National Science Foundation.}). Wavelength calibration was achieved with Helium and Argon arc lamps.

\subsection{Spectroscopy from AAO}

The AAT spectra were obtained with the fibre-fed dual-beam AAOmega spectrograph on the 3.9\,m Anglo-Australian Telescope (AAT) at the Australian Astronomical Observatory on 2012, July 7--8. The Two Degree Field (``2dF") multi-object system was utilised. Light from an optical fibre of diameter $2\farcs1$ on the sky is fed into two arms via a dichroic beam-splitter with crossover at 5\,700\AA. Each arm of the AAOmega system is equipped with a 2k$\times$4k E2V CCD detector and an AAO2 CCD controller. The blue arm CCD is thinned for improved blue response. Because of the atmospheric diffraction, our targets did not produce useful spectra on the red arm, which was observing around the infrared Ca\,{\small II} triplet for a different programme. We used grating 580V, giving $R=1\,300$ over $\sim2100$\AA. The central wavelength was set at 4\,500\AA.

We used the standard reduction pipeline {\tt 2dfdr} as provided by the AAT at the time, with wavelength calibration by observing arc lamps before each target exposure. The arc lamps provide lines of He+CuAr+FeAr+ThAr+CuNe, and only those lines actually detected in the calibration exposures were input into the data reduction pipeline. The wavelength calibration was excellent, with rms consistently $<0.1$ pixel.

Sky subtraction was carried out by means of a mean sky spectrum, obtained by averaging the spectra of 30 fibres positioned at known blank locations. The sky lines in each spectrum are evaluated and used to scale the mean sky spectrum prior to subtraction.

The journal of the spectroscopic observations is presented in Table~\ref{TabObs}.

\begin{table*}
\centering
\caption{Summary of observations. The first column `Campaign' is used to identify the setup for each individual source listed in Table~\ref{TabResults}.}
\label{TabObs}
\begin{tabular}{lllll}
\hline
\noalign{\smallskip}
Campaign & Telescope & Dates & Wavelength range & Resolution\\
 & & & (\AA\AA) & (\AA) \\
\hline\hline
eso07 & ESO 3.6\,m, La Silla, Chile & 2007 Sep 19--20  & 3082--5087  & 4  \\
eso11 & ESO NTT, La Silla, Chile & 2011 Dec 9--10& 3095--5085 & 10 \\
aat & AAT, Siding Spring, Australia & 2012 Jul 7--8 & 4025--4775 & 1.2 \\
\hline
\end{tabular}

\end{table*}

\subsection{Photometry from OGLE}
\begin{figure*}
\includegraphics[width=1.0\textwidth]{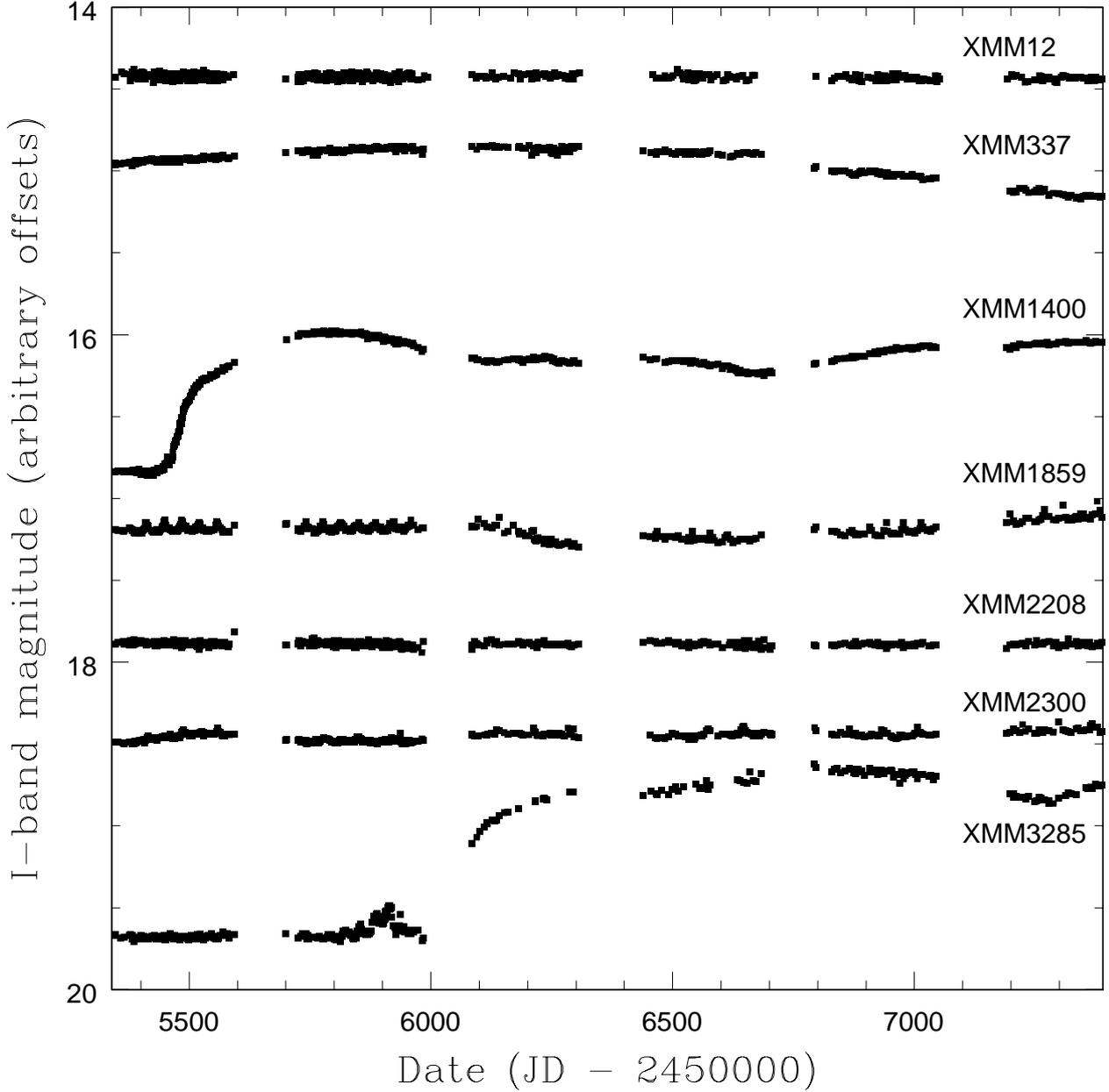}
\caption{OGLE IV lightcurves of our six HMXB candidates and XMM 2300 to illustrate the extent of variability in the lightcurves.}
\label{ogle_all}
\end{figure*}

The OGLE project\footnote{http://ogle.astrouw.edu.pl} (see e.g. \citealt{UdalskiKubiakSzymanski1997},  \citealt{UdalskiSzymanskiSzymanski2015}) provides long term $I$-band photometry with roughly daily sampling. These data were used to study the time variability of the optical counterparts to the eight candidate HMXBs.  All of the HMXB candidates have been observed with OGLE III \& IV. The OGLE identifications and lightcurve characteristics are presented in Table~\ref{ogletab} and the OGLE IV lightcurves in Fig.~\ref{ogle_all}.

\section{Analysis}
\label{Method}

\subsection{Criteria for identification as a massive X-ray binary}
Starting with the list of massive X-ray binary candidates in Table 5 of \citet{SturmHaberlPietsch2013}, and excluding any objects previously confirmed as X-ray binary systems, we applied the following criteria:
\begin{enumerate}
\item the X-ray source must belong to confidence classes 2 or 3 -- description below
\item the optical counterpart must be a spectroscopically confirmed early-type star
\end{enumerate}

The confidence classes listed above were introduced by \citet{HaberlSturm2016} in their classification of X-ray binaries in the SMC. {\bf Class 2} corresponds to an X-ray object showing significant X-ray variability ($\ge$ a factor of 30) or a hard power law spectrum, but no X-ray pulsations. {\bf Class 3} corresponds to an X-ray source with an error circle small enough to identify it unambiguously with an early type star showing emission lines. 

Though we do not require a detected period in the optical lightcurve to classify an object as an HMXB, some systems do exhibit strong evidence for binary modulation. However, we also note that many HMXB systems do not exhibit this modulation \citep{BirdCoeMcBride2012}. As a result of the above selection criteria we present here {\bf six} new HMXB systems.

\subsection{Spectral classification}

\begin{figure}
\includegraphics[width=0.52\textwidth]{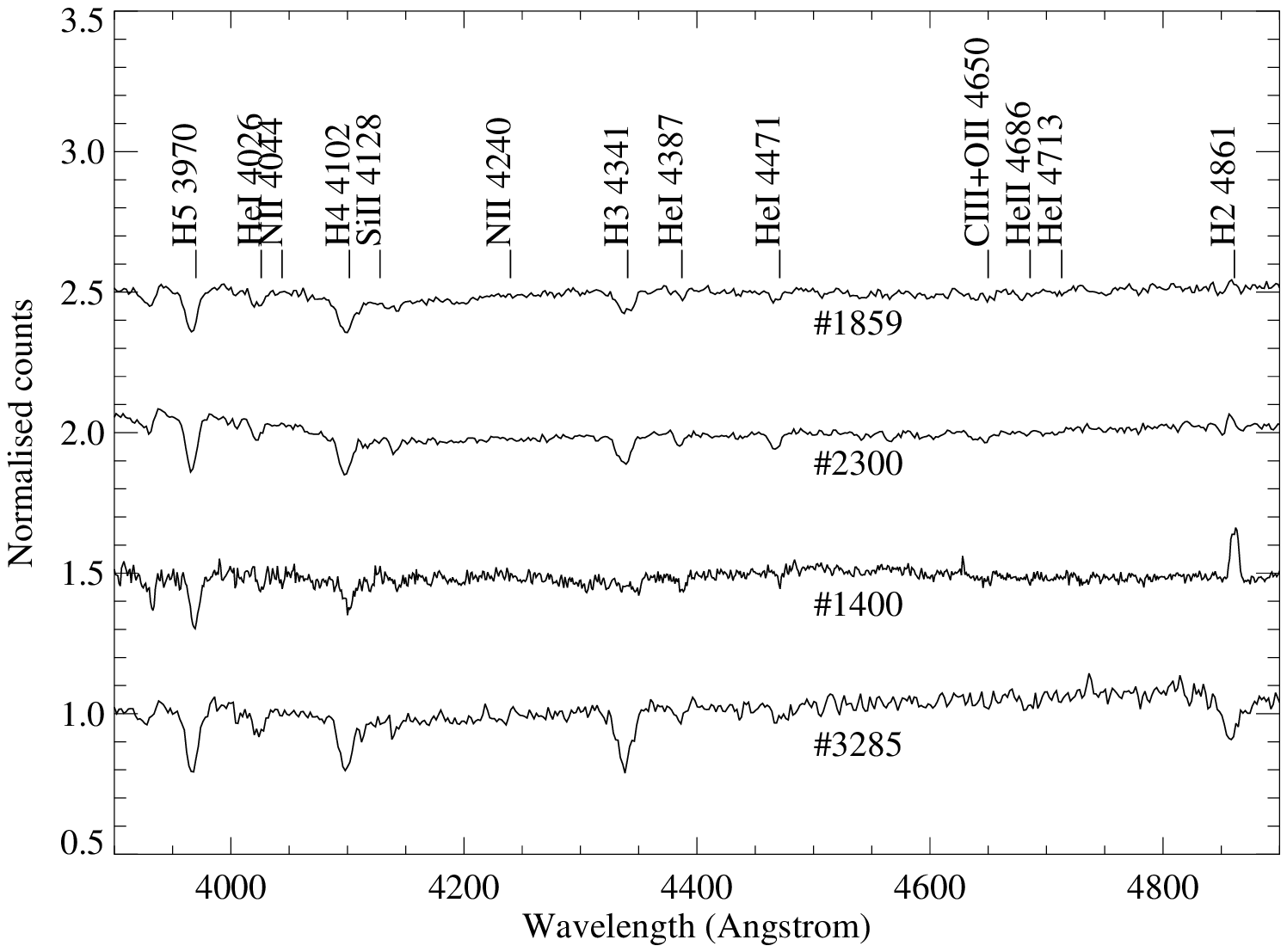}
\includegraphics[width=0.52\textwidth]{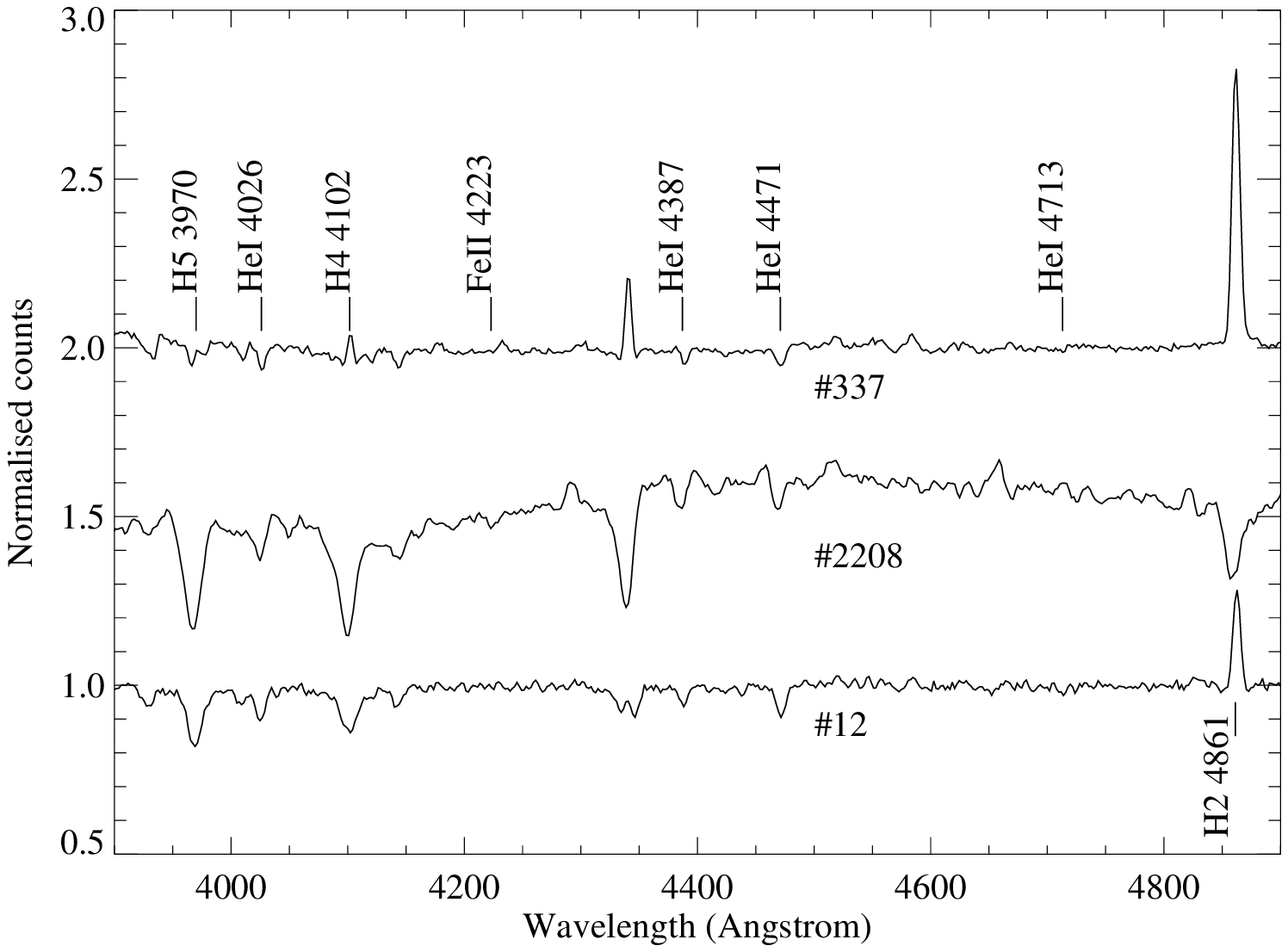}
\caption{The top panel shows those spectra classified as B0 -- B1 in this work. The lower panel shows those classified as later than B1.}
\label{FigSpec}
\end{figure}

The motivation for and method of spectral classification has been outlined in section 4 of \citet{McBrideCoeNegueruela2008}.
We show in Fig.~\ref{FigSpec} the spectra we classified in this work. Spectral classifications can be found in Table~\ref{TabResults}.

\subsection{Timing analysis}
The OGLE data sets used in this analysis are listed in Table~\ref{ogletab} and provide $I$-band photometry over many years.

\begin{table*}
\caption{OGLE III \& IV identifications and variability parameters. $\delta$I is the mean error on the data points, whereas $\Delta$I represents the total range of the data set.}
\vbox {\vfil
\begin{tabular}{llllll}
  \hline
        XMM&   I band         & $\delta$I & $\Delta$I & OGLE III & OGLE IV \\
           ID& average mag   & (mmag) & (mmag) & ID& ID \\
 \hline
 12 & 15.53 &5&105&SMC121.8 55&SMC733.20 2992\\
 337 &14.42 &3&682&smc105.6.39454 &SMC726.32 26912 \\
 1400 & 14.90&4&955&smc100.4.62974&smc719.03.43645\\
 1859 & 14.82&3&867&smc103.7.6594&SMC720.11 13342\\
 2208 & 16.92&7&273&SMC108.5 8790 &SMC719.26 20124\\
 2300 & 14.45&3&178&SMC105.5 37304&SMC719.18 7\\
 3285 & 15.70&5&1246&SMC110.5 9503&SMC726.28 23178\\

\hline
\end{tabular}
\vfil}
\label{ogletab}
\end{table*}

The optical lightcurves were subject to detrending and period searching following the method explained in sections  2 and 3 of \citet{BirdCoeMcBride2012}. The results of this timing analysis are presented in Table~\ref{TabResults}.

\begin{table*}
  \caption{Results from optical spectroscopy and OGLE data timing analysis. Note there are two periods seen in source XMM 12, and that the periodicity seen in source No. 3285 is only visible in one year's worth of data covering the period MJD 55650--56100.}
  \label{TabResults}
  \begin{tabular}{lllllllll}
  \hline
   XMM     &   RA         & Dec & V mag & Campaign & Sp Type & Period  & Detrending & Previously referenced\\
         ID    & (J2000)   & (J2000) & MCPS & & this work & (days) & (days) & \\
 \hline
 12 & 01 19 38.94  & -73 30 11.4 & 15.8 & eso07 &  B1.5--B2\,III-Ve & 0.5522541(18)  & 81& [SG05]28, [MA93]1867,\\
  &  &  &  & &  & 5.18331(24)  & 81& [HS00]60\\
 337 & 00 56 14.65  & -72 37 55.8& 14.6 & eso07 & B3\,IIIe & - & 81 & [SG05]22, [MA93]922,\\
  & & & & & or earlier& & &[E04]1135\\
1400 & 00 53 41.76 & -72 53 10.1& 14.7 & aat & B0.5--B1\,IIIe & - & 81 & \citet{LambOeyGraus2013}\\
1859 & 00 48 55.55 & -73 49 46.4& 14.9 & eso11 & B0\,IV-Ve & 36.432(9) & 81 & [E04]705, \\ 
  & & & & & & & &\citet{LambOeySegura2016}\\
  & & & & & & & &\citet{CoeMcBrideHaberl2016}\\
2208 & 00 56 05.48 & -72 00 11.1& 16.7 & eso11& B2--B3\,Ve & - & 81 & \citet{NovaraLaPalombaraMereghetti2011}\\
2300$\dagger$& 00 56 13.87 & -72 29 59.7& 14.5 & eso11+aat& B0.5\,IVe& 72.231(35) & 101 & \citet{EvansLennonSmartt2006}\\
3285 & 01 04 29.42 & -72 31 36.5& 15.8 & eso11+aat & B1\,V& 29.75(18) & 41 & \citet{RajoelimananaCharlesUdalski2011}, \\
 & & & & & & & &\citet{SchmidtkeCowleyUdalski2013} \\

 \hline
\end{tabular}
 {\small MCPS: Magellanic Clouds Photometric Survey \citealt{ZaritskyHarrisThompson2002}, SG05: \citealt{ShtykovskiyGilfanov2005}, MA93: \citealt{MeyssonnierAzzopardi1993}, HS00: \citealt{HaberlSasaki2000} E04: \citealt{EvansHowarthIrwin2004} $\dagger$: This source is a previously known HMXB \citep{SturmHaberlPietsch2013} and we report here our spectral classification and periodicity.}
\end{table*}

\section{Discussion}
\label{Results}

\subsection{Spectroscopy results}

In Table~\ref{TabResults} we present spectral classifications of the optical counterparts to seven X-ray sources. Six of the candidate X-ray binaries are now confirmed HMXBs, while XMM 2300 is identified already in \citet{SturmHaberlPietsch2013} as an X-ray source corresponding to a Be star\citep{EvansLennonSmartt2006}. In all cases the counterpart is classified as a B star, confirming the HMXB nature of the system and validating the X-ray and optical selection criteria as used by \citet{SturmHaberlPietsch2013}. The spectra are illustrated in Fig.~\ref{FigSpec}. Our spectral type of XMM 337 is confirmed by \citet{EvansHowarthIrwin2004}, but the same authors find a slightly earlier spectral type (O9.5\,V, their object \#705) for XMM 1859, while \citet{LambOeySegura2016} find a spectral type of O9\,IIIe for XMM 1859. For XMM 1400, \citet{LambOeyGraus2013} find a spectral type of B0\,Ve. Our classification of XMM 2300 confirms that of \citet{EvansLennonSmartt2006}.

Two spectra, XMM 2208 and XMM 3285 do not show emission in the H$\beta$ line. XMM 2208 has shown strong H$\alpha$ in emission in previous observations \citep{NovaraLaPalombaraMereghetti2011}.  
 Be stars are known to undergo disk loss phases, where the Balmer lines appear entirely in absorption, so despite the lack of emission lines, the optical period, $V$ magnitude and shallow H$\beta$ absorption line (compared with the other Balmer lines in this source) all point strongly towards a Be X-ray binary nature of XMM 3285. 

\subsection{Timing results}
\citealt{BirdCoeMcBride2012} show that there can be confusion between short period ($\le 1$ day) non-radial pulsations and longer period ($>10$ days) orbital modulation when the approximately daily sampling period beats with the non-radial pulsation (NRP) period and results in long period peaks in the periodogram.  A case in point is XMM 3285, where a 37.15\,d period reported by \citet{RajoelimananaCharlesUdalski2011} was associated with aliasing of a possible shorter period modulation \citep{SchmidtkeCowleyUdalski2013}.

\begin{figure*}
\includegraphics[width=0.6\textwidth, angle=270]{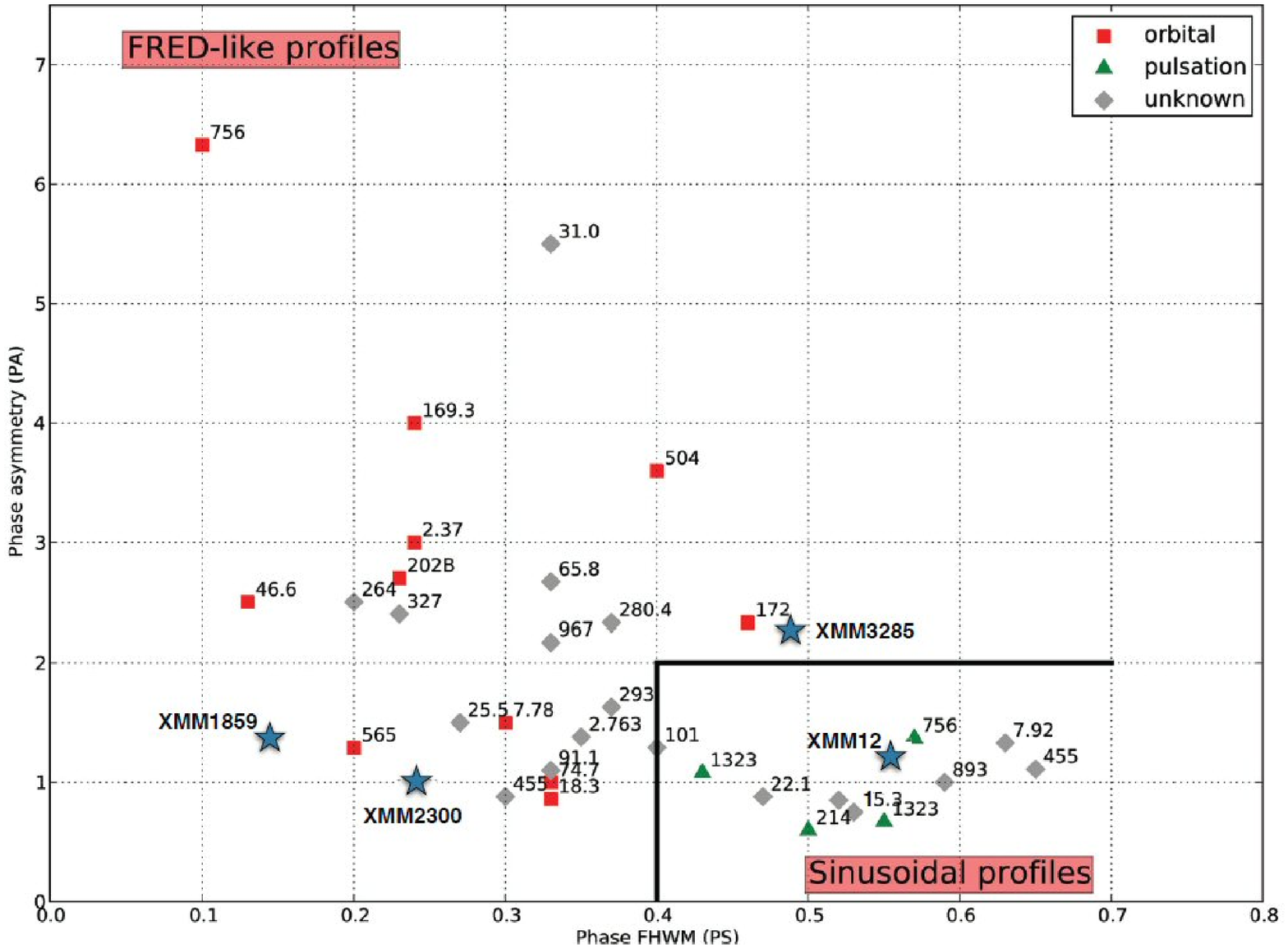}
\caption{The parameters space for discriminating between aliased non-radial pulsations and orbital modulations in terms of the metrics phase asymmetry and phase FWHM \citep{BirdCoeMcBride2012}. This figure shows the positions of the four sources exhibiting periodicities from this work against the background of all the previous sources studied in this way by \citet{BirdCoeMcBride2012}. Previous sources are all confirmed Be X-ray binaries and are labelled by their neutron star spin periods (in seconds).}
\label{FigFred}
\end{figure*}

An effective discriminant between aliased non-radial pulsations and orbital modulation is the shape of the folded lightcurve.  The orbital modulation, probably caused by the neutron star disturbing the Be star circumstellar disk typically shows a fast rise with an exponential decay, whereas non-radial pulsations have more sinusoidal profiles in the lightcurves.  These characteristics can be reflected by two metrics: the phase asymmetry of the folded lightcurves, and the phase FWHM \citep{BirdCoeMcBride2012} with the parameter space shown in Fig.~\ref{FigFred}.

From this plot it can be seen that only sources XMM 1859 \& XMM 2300 show strong evidence for the presence of a modulation that seems incompatible with the sinusoidal behaviour associated with NRP behaviour. Consequently we present these results as clear evidence for binary modulation. Source XMM 3285 is marginal, whereas XMM 12 is a clear NRP candidate. XMM 12 exhibits two periods, at 0.55\,d and at 5.18\,d. Both periods are plotted, and lie on top of each other, in Fig.~\ref{FigFred}.

\section{Concluding Remarks}
\label{Conclude}

In this work we have presented an optical spectroscopic and timing analysis of candidate high mass X-ray binaries in the Small Magellanic Cloud that generally exhibit low X-ray luminosities. The observations presented confirm the Be X-ray binary nature of these six candidates. We find evidence of two probable binary periods: 36.4\,d in XMM\,1859 and 72.2\,d in XMM\,2300. All candidates have early spectral types, as expected from the selection criteria applied in the original candidate HMXB list by \citet{SturmHaberlPietsch2013}.

Recently a spin period of 15.6\,s was discovered in XMM 1859 \citep{VasilopoulosHaberlAntoniou2016}, illustrating that the Be X-ray binaries as selected by this sample have properties consistent with those of the general population of Be X-ray binaries in the SMC. It is likely that they've been previously undetected as they've never been observed during an X-ray bright phase. Swift has just started regular monitoring of the SMC \citep{EvansKennaCoe2016}, and this improved sensitivity together with wide area coverage may unearth more Be X-ray binary systems that have eluded detection up until now.

\section*{Acknowledgments}
The AAT observations have been supported by the OPTICON project (observing proposals 2011A/014 and 2012/A015), which is funded by the European Commission under the Seventh Framework Programme (FP7).
VAM acknowledges financial support from the National Research Foundation of South Africa (Grant RDYR14081189466) and the World Universities Network. RD, AM, IN from the University of Alicante acknowledge support from the Spanish Government Ministerio de Econom\'{\i}a y Competitividad under grant  AYA2015-68012-C2-2-P (MINECO/FEDER). ESB acknowledges support from a Claude Leon Foundation fellowship and from the Marie Curie Actions of the European Commission (FP7-COFUND). The OGLE project has received funding from the National Science Centre, Poland, grant MAESTRO 2014/14/A/ST9/00121 to AU.

\bibliographystyle{mnras}


\end{document}